\documentclass{article}
\usepackage{spconf,amsmath,graphicx}
\usepackage{subfig}
\usepackage{algorithm}
\usepackage[noend]{algorithmic}
\usepackage{listings}
\usepackage{booktabs}
\usepackage{url}
\usepackage{hyperref}

\title{Label-Looping: Highly Efficient Decoding for Transducers}
\name{Vladimir Bataev\textsuperscript{1,2*}, Hainan Xu\textsuperscript{1*}, Daniel Galvez\textsuperscript{1}, Vitaly Lavrukhin\textsuperscript{1}, Boris Ginsburg\textsuperscript{1}}
\address{
$^1$NVIDIA \ \ \
$^2$University of London, UK
}

\begin{document}
%
\maketitle
\begingroup\renewcommand\thefootnote{*}
\footnotetext{Equal contribution}
\endgroup
\begin{abstract}
This paper introduces a highly efficient greedy decoding algorithm for Transducer-based speech recognition models. We redesign the standard nested-loop design for RNN-T decoding, swapping loops over frames and labels: the outer loop iterates over labels, while the inner loop iterates over frames searching for the next non-blank symbol. Additionally, we represent partial hypotheses in a special structure using CUDA tensors, supporting parallelized hypotheses manipulations. Experiments show that the label-looping algorithm is up to 2.0X faster than conventional batched decoding when using batch size 32. It can be further combined with other compiler or GPU call-related techniques to achieve even more speedup. Our algorithm is general-purpose and can work with both conventional Transducers and Token-and-Duration Transducers. We open-source our implementation to benefit the research community.
\end{abstract}

\section{Introduction}
\label{sec:intro}

Connectionist temporal classification (CTC)\cite{graves2006connectionist} and Transducer \cite{graves2012sequence} are the most popular architectures for end-to-end automatic speech recognition (ASR). Both CTC and Transducer adopt a frame-synchronous paradigm. Both have a special blank symbol as an output label so that the model output can be of a different length than the audio input. The major difference between those models is that CTC adopts a conditional independence assumption in predicting tokens from each frame. At the same time, Transducers consider textual context during token prediction, making it achieve better accuracy than CTC models at a slightly increased computational cost.

Many research works have attempted to improve the efficiency of Transducer models.
More powerful encoder architectures, e.g., Transformers \cite{vaswani2017attention} and Conformers \cite{gulati2020conformer} are proposed to replace the original LSTM encoders \cite{graves2012sequence}, which both improve the model's accuracy and efficiency.
\cite{ghodsi2020rnn} replaced the LSTM predictor with a stateless network, which shows a speed-up of running Transducer models with a slight degradation in model performance. 
\emph{Multi-blank Transducers}~\cite{xu2022multi} introduced \emph{big blank} tokens that allow skipping of multiple frames during decoding when blank symbols are predicted during inference, bringing significant speed-up and also slightly improving model accuracy;
\emph{Token-and-Duration Transducer} (TDT)~\cite{xu2023efficient} models extend the multi-blank model by decoupling the prediction of tokens and durations so that the model can skip multiple frames after each prediction, regardless of whether it's a blank, bringing even more speed-up to Transducer inference. 
\cite{kang2023fast} added additional constraints to Transducers in terms of the maximum labels emitted per frame, which shows speed-up for its inference. \cite{tian2021fsr} proposed to use a CTC model to help predict blank symbols in advance so the overall decoding time of the model can be reduced. 
\cite{saon2020alignment,kim2020accelerating,kim2020accelerating2} all proposed alternative ways to expand search paths to speed up beam search for Transducers.

This paper focuses on improving the greedy decoding algorithm for Transducer models since greedy search provides the best trade-off between speed and accuracy when the model is well-trained on a sufficient amount of domain data~\cite{prabhavalkar2023survey}.
We propose a \emph{label-looping} algorithm for batched inference, which achieves significant speed-up. 
The contribution of this paper is as follows:
\begin{enumerate}
    \item A novel \emph{label-looping} algorithm that separates the processing of blank and non-blank emissions of Transducer decoding, maximizing parallelism.
    \item A data structure to represent partial hypotheses within each batch that can be efficiently manipulated using PyTorch \cite{paszke2019pytorch} CUDA tensor operations.
   \item Label-looping algorithm is a general-purpose algorithm applicable to both Transducer and TDT models and supports stateful (LSTM) and stateless prediction networks. This is also the first efficient implementation of exact batched decoding for TDT.

\end{enumerate}
The proposed algorithm brings up to 2.0X speedup compared to existing batched decoding algorithms. The speedup can be increased up to 3.2X when combined with technique in compilation and GPU call optimization. 
The label-looping decoding is open-sourced in the NeMo~\cite{kuchaiev2019nemo} toolkit\footnote{\href{https://github.com/NVIDIA/NeMo/blob/r2.0.0/nemo/collections/asr/parts/submodules/rnnt_loop_labels_computer.py}{\begin{scriptsize}\texttt{github.com/NVIDIA/NeMo/.../rnnt\_loop\_labels\_computer.py}\end{scriptsize}}}\footnote{\href{https://github.com/NVIDIA/NeMo/blob/r2.0.0/nemo/collections/asr/parts/submodules/tdt_loop_labels_computer.py}{\begin{scriptsize}\texttt{github.com/NVIDIA/NeMo/.../tdt\_loop\_labels\_computer.py}\end{scriptsize}}}.

\section{Background}

\begin{figure}[t]
    \centering
    \includegraphics[width=5.98cm]{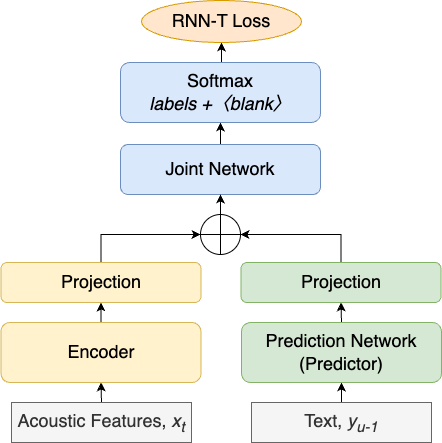}
    \caption{Transducer Architecture}
    \label{fig:rnnt}
\end{figure}

The Transducer model is a popular end-to-end ASR architecture~\cite{graves2012sequence}. As shown in Fig.~\ref{fig:rnnt}, it consists of an encoder, a prediction network (also called predictor), and a joiner. The encoder and predictor extract higher-level information from the audio and the text history context, respectively, and the joiner combines the outputs of the encoder and prediction network and computes a probability distribution over the vocabulary. The Transducer model has a special $\langle blank \rangle$ (also denoted as $\langle b \rangle$) symbol in its vocabulary, which serves as a delimiter between labels (zero or more) corresponding to different frames.

The inference of the Transducer model requires traversing the encoder output frame by frame and iteratively predicting tokens based on the prediction history. Algorithm \ref{RNNT_algo} shows the inference procedure of the Transducer model on a single audio utterance, where the blank symbol is represented as $\langle b \rangle$. Note that 
(lines 11 to 15): 
\begin{itemize}
    \item $t$ is incremented only after $\langle blank \rangle$ predictions and remains unchanged for non-blank predictions.
    \item only non-blank predictions are added to the output and used to update predictor states.
\end{itemize}
\begin{algorithm}[t]
   \caption{Inference of  Transducer}
   \label{RNNT_algo}
\begin{algorithmic}[1]
   \STATE {\bfseries input:} acoustic input $x$
    \STATE enc = encoder(x) \textit{\# output dim is [T, dim]}
    \STATE hyp, state, $t$ = [], predictor.init\_state(), 0
    \WHILE{$t <$ T}
    \IF {hyp is []}
        \STATE dec, new\_state = predictor(state, BOS)
    \ELSE
        \STATE dec, new\_state = predictor(state, hyp[-1])
    \ENDIF
    \STATE token\_probs = joiner(enc[$t$], dec)
    \STATE prediction = argmax(token\_probs)
    \IF{prediction $\neq \langle b \rangle$ }
    \STATE hyp.append(prediction)
    \STATE state = new\_state
    \ELSE
    \STATE $t$ = $t$ + 1 
    \ENDIF
    \ENDWHILE
    \STATE {\bfseries return} hyp 
\end{algorithmic}
\end{algorithm}

\begin{algorithm}[t]
   \caption{Batched Inference of Transducer}
   \label{RNNT_batch_algo}
\begin{algorithmic}[1]
   \STATE {\bfseries input:} acoustic input $x_1, x_2, ..., x_B$
    \STATE encs = encoder($x$)  \# output dim is [B, T, dim]
    \STATE hyps, states, t = [[] * B], [predictor.init\_state() * B], 0
    \STATE predictions = [BOS * B]
    \WHILE{$t <$ T}
        \STATE decs, new\_states = predictor(states, predictions)
        \STATE token\_probs = joiner(encs[$t$], decs)
        \STATE predictions = argmax(token\_probs)
        \STATE blank\_mask = (predictions $== \langle b \rangle$)
        \STATE states[\textbf{not} blank\_mask] = new\_states[\textbf{not} blank\_mask]
        \WHILE{\textbf{not} blank\_mask.all()}
        \FOR{i = 1 to B}
        \IF{\textbf{not} blank\_mask[i]}
        \STATE hyps[i].append(predictions[i])
        \ENDIF
        \ENDFOR
        \STATE decs, new\_states = predictor(states, predictions)
        \STATE token\_probs = joiner(encs[$t$], decs) 
        \STATE predictions = argmax(token\_probs)
        \STATE blank\_mask = blank\_mask $||$ (predictions $== \langle b \rangle$)
        \STATE states[\textbf{not} blank\_mask] = \textbackslash \newline
            \hspace*{5em}new\_states[\textbf{not} blank\_mask]
        \ENDWHILE
        \STATE $t = t + 1$
    \ENDWHILE
    \STATE {\bfseries return} hyps
\end{algorithmic}
\end{algorithm}

The difference in processing blank and non-blank predictions presents difficulties for efficient batch inference for Transducers: the time index $t$ for different utterances can be incremented at different times, and partial hypotheses for different utterances might grow asynchronously as well. This makes it hard to implement an efficient batched inference algorithm that can fully leverage parallel computing.
A common implementation of batched inference for Transducers, as supported in open-source toolkits like ESPNet~\cite{watanabe2018espnet}, NeMo~\cite{kuchaiev2019nemo} and SpeechBrain~\cite{ravanelli2021speechbrain}, 
is shown in Algorithm \ref{RNNT_batch_algo} (for simplicity, this Algorithm assumes all audio in the batch has the same length $T$; in practice, we need extra checks in the code for individual audio lengths). This algorithm has the following limitations in terms of efficiency:
\begin{enumerate}
    \item The algorithm increments the time stamp $t$ for different utterances synchronously, and if one utterance predicts a non-blank symbol, all other utterances in the same batch must wait before the whole batch advances to the next time step.
    \item Lines 10 and 19 selectively update the predictor states and outputs, depending on whether the last prediction is blank. This means some of the predictor state computation is wasted.
    \item At lines 12 to 14, processing of different hypotheses in the batch is done sequentially with a for-loop to handle blank and non-blank predictions differently, which can be time-consuming.
\end{enumerate}

\section{Label-looping Decoding Algorithm}

\begin{algorithm}[t]
   \caption{Label-looping Algorithm}
   \label{RNNT_label_loop_algo}
\begin{algorithmic}[1]
   \STATE {\bfseries input:} acoustic input $x_1, x_2, ..., x_B$, input\_length
    \STATE encs = encoder($x$)   \# output dim is [B, T, dim]
    \STATE hyps, state = [[] * B], [predictor.init\_state() * B]
    \STATE b2active, b2time = [True * B],  [0 * B]
    
    \WHILE{b2active.any()}
    \STATE decs, states = predictor(state, predictions)
    \STATE token\_probs = joiner(encs[b2time], decs)
    \STATE predictions = argmax(token\_probs)
    \STATE blank\_mask = (predictions $== \langle b \rangle$)
    \STATE b2time[blank\_mask] += 1
    \STATE b2active = b2time $\leq$ input\_length
    \WHILE{(blank\_mask \textbf{AND} b2active).any()}
        \STATE token\_probs = joiner(encs[b2time], decs)
        \STATE extra\_predictions = argmax(token\_probs)
        \STATE extra\_blank\_mask = (predictions $== \langle b \rangle$)
        \STATE predictions[blank\_mask] = \textbackslash \newline
            \hspace*{5em}extra\_predictions[blank\_mask]
        \STATE blank\_mask = blank\_mask \textbf{AND} extra\_blank\_mask
        \STATE b2time[blank\_mask] += 1
        \STATE b2active = b2time $\leq$ input\_length
    \ENDWHILE
    \STATE hyps.append(predictions)
    \ENDWHILE
    \STATE {\bfseries return} hyps 
\end{algorithmic}
\end{algorithm}

\begin{figure}[ht!]
        \subfloat[\centering Conventional (frame-looping) decoding algorithm]{{
            \includegraphics[width=7.6cm]{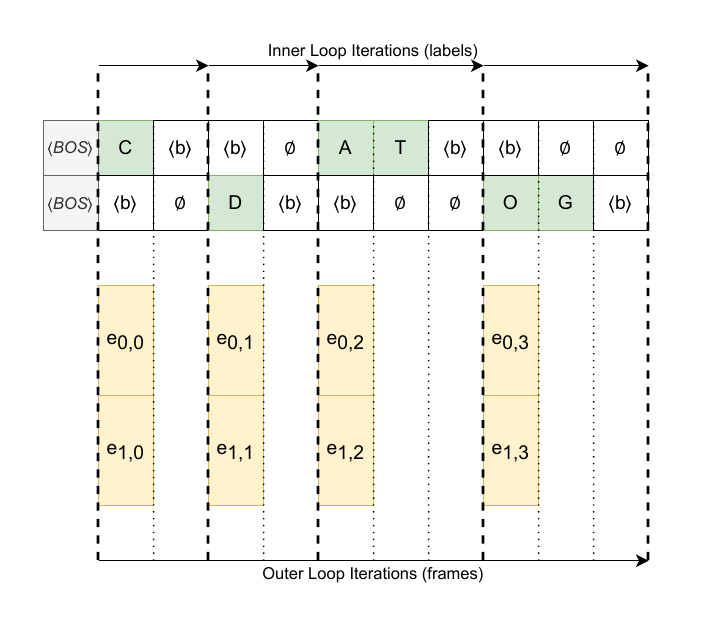}
            \label{fig:decoding-algorithms-frame-looping}
        }}
        \qquad
        \subfloat[\centering Label-looping algorithm - proposed]{{
            \includegraphics[width=4.6cm]{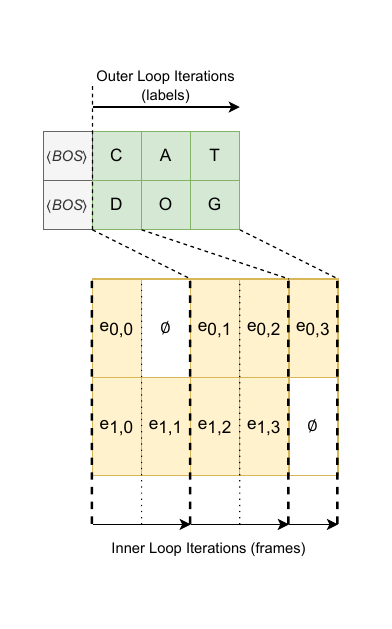}
            \label{fig:decoding-algorithms-label-looping}
        }}
	\hfill
	\caption{\emph{Frame-looping} and \emph{label-looping} decoding algorithms operations. The batch size is 2, and the length of encoder output is 4. \newline Ground truth transcriptions are ``CAT" and ``DOG". \newline Alignments: `C $\langle b \rangle$ $\langle b \rangle$ A T $\langle b \rangle$ $\langle b \rangle$', `$\langle b \rangle$ D $\langle b \rangle$ $\langle b \rangle$ O G $\langle b \rangle$'. $\emptyset$ symbol indicates unnecessary computations in the algorithm due to batched decoding.}
	\label{fig:decoding-algorithms}
\end{figure}

The label-looping method for batched Transducer inference consists of two novel components:
\begin{enumerate}
    \item an algorithm to run inference on Transducers that maximizes the parallelism, and
    \item a data structure to represent hypotheses in a batch that supports parallelized manipulation of hypotheses.
\end{enumerate}

\subsection{Representation of batched hypotheses}
To represent partial hypotheses during inference, instead of having a single \emph{Hypothesis} class that stores the tokens, time-stamps, scores, and other important elements, we store information about hypotheses in a batch in one \emph{BatchedHyps} class, in which all information is stored with CUDA tensors. Operations like adding tokens to partial hypotheses, updating scores, and time-stamps are implemented with masked CUDA tensor operations for both correctness and high efficiency.
Our design stores the texts of partial hypotheses with a 2D tensor of [B, max-length], where B is the number of utterances in the batch, and the max-length is initialized with a smaller value proportional to the length of the audio and will double the size if this max-length is exceeded.
We show a simple pseudo-code for the class below and would advise readers to check our open-sourced implementation for details\footnote{\href{https://github.com/NVIDIA/NeMo/blob/r2.0.0/nemo/collections/asr/parts/utils/rnnt_utils.py}{\begin{scriptsize}\texttt{github.com/NVIDIA/NeMo/.../rnnt\_utils.py}\end{scriptsize}}}. 

\lstset{basicstyle=\ttfamily}

\begin{lstlisting}[language=Python]
class BatchedHyps:
  current_lengths: LongTensor[B]
  transcripts: LongTensor[B, max_length]

  def add_results(self, add_mask, labels):
    self.transcripts[range(B), 
      self.current_lengths] = labels
    self.current_lengths += add_mask
\end{lstlisting}

\subsection{Label-looping algorithm for RNN-Transducers}

The proposed algorithm is shown in Algorithm \ref{RNNT_label_loop_algo}. It is based on the observation that only \emph{non-blank predictions} require prediction network state updates, so we can maximize the parallelism of computation by having the outer loop (line 5) of the algorithm process only \emph{non-blank predictions}. The less expensive \emph{blank predictions} are processed in the inner loop (line 12). With this design, after the prediction network is executed (line 6), there is no need to selectively keep the updated states, a process that is required in the original Algorithm \ref{RNNT_batch_algo} (lines 10 and 19).

The difference between conventional (frame-looping) and label-looping approaches is visualized in Fig.~\ref{fig:decoding-algorithms}. When working with batch size larger than 1, a conventional decoding algorithm (see Fig.~\ref{fig:decoding-algorithms-frame-looping}) can find non-blank labels in utterances at different steps. Each inner loop iteration requires a prediction network update. Although it is possible to evaluate the prediction network only after finding new labels (indicated by green color), and variable batch size can be used to evaluate it only for necessary elements, we found that such techniques do not lead to significant speed improvements. In the label-looping approach (see Fig.~\ref{fig:decoding-algorithms-label-looping}), the number of prediction network calls is equal to the length of the largest hypothesis, which is the least possible number. This not only allows the running of existing models efficiently but also minimizes the negative performance impact of potentially larger prediction network. Also, we found that further optimization of lightweight joint network calls to avoid all possible extra computations does not improve performance after minimizing operations, as described in Sec.~\ref{sec:precompute-projections}.

\subsection{Label-looping for Token-and-Duration Transducers}
Token-and-duration Transducers (TDT) \cite{xu2023efficient} is an extension of Transducer models that decouples token and duration prediction. With TDT, $t$ can be updated regardless of whether the predicted token is $\langle blank \rangle$ and can be incremented by more than 1.
The label-looping algorithm for the TDT model is mostly similar to that of conventional Transducers. The only difference is at lines 10 and 18, when updating \emph{b2time}, the increment amount of the TDT model comes from the model's duration prediction. \footnote{Due to space limitations, we refer our readers to our open-source implementation for the actual algorithm.}

\subsection{Precomputation of encoder/predictor projections}
\label{sec:precompute-projections}

In the standard Transducer model architecture, two linear layers are required to project the encoder and prediction network outputs into the same vector space. Since this computation is repeatedly run through the decoding process,  it can become a significant overhead. In our algorithm, we propose precomputing those projections, at lines 2 and 6 in Algorithm~\ref{RNNT_label_loop_algo}, before feeding them to the joiner.

We point out that it is also possible to precompute the projections with the original Algorithm~\ref{RNNT_batch_algo} (lines 2, 6, and 15). However,
the predictor updates (lines 6 and 15) happen much more frequently, and part of the output would be filtered out at lines 10 and 17. As a result, the precomputation of projections would have a smaller impact on the original algorithm.

\section{Experiments}

We compared our label-looping algorithm against the baseline algorithm based on the Algorithm~\ref{RNNT_batch_algo} available publicly in the NeMo~\cite{kuchaiev2019nemo} toolkit.
We used two RNNT models and two TDT models. RNNT-L\footnote{\href{https://hf.co/nvidia/stt_en_fastconformer_transducer_large}{\begin{scriptsize}\texttt{hf.co/nvidia/stt\_en\_fastconformer\_transducer\_large}\end{scriptsize}}} 
and TDT-L\footnote{Due to legal requirements, this model is not open-sourced yet at the time of publication. We plan to retrain the model, open-source it, and update the results accordingly. The ckeckpoint is planned to be released at \href{https://hf.co/nvidia/stt_en_fastconformer_tdt_large}{\begin{scriptsize}\texttt{hf.co/nvidia/stt\_en\_fastconformer\_tdt\_large}\end{scriptsize}}.}
(Large) have 114M parameters, 
and RNNT-XXL\footnote{\href{https://hf.co/nvidia/parakeet-rnnt-1.1b}{\begin{scriptsize}\texttt{hf.co/nvidia/parakeet-rnnt-1.1b}\end{scriptsize}}} 
and TDT-XXL\footnote{\href{https://hf.co/nvidia/parakeet-tdt-1.1b}{\begin{scriptsize}\texttt{hf.co/nvidia/parakeet-tdt-1.1b}\end{scriptsize}}} have 1.1B parameters.

Both models use Fast-Conformer~\cite{rekesh2023fast} encoder with 8X subsampling and 1024 BPE~\cite{sennrich2015neural} vocabulary size at the output side.  
For each model, we report its \emph{inversed real-time factor}~\footnote{defined as the ratio of audio length to its decoding time. E.g., if it takes 1 second to decode 2-second audio, then RTFx = 2.} (RTFx) on the LibriSpeech \cite{panayotov2015librispeech} test-other dataset. To provide a better picture, we also include the RTFx for time spent in decoding only, excluding encoder computations, since the latter can more directly reflect the speed-up brought by our algorithms. To get a more accurate measurement, we first run decoding twice to ``warm-up'' the cache required for those algorithms to work more efficiently and report the average time from the third to fifth measurement. All experiments are done on NVIDIA RTX A6000 GPU with bfoat16 precision.

\subsection{Results with RNN Transducers}

\begin{table}[t]
    \centering
    \caption{Transducer (RNNT) decoding speed of different algorithms and batch-sizes on LibriSpeech test-other. We report ``total computation RTFx / non-encoder RTFx'', where ``RTFx" is the inversed real-time factor. WER is 3.9\% for the RNNT-L model and 2.7\% for the RNNT-XXL model.
    }
    \begin{tabular}{l c c c c}
    \toprule
    batch    & RTFx, baseline & RTFx, label-looping & rel. speedup  \\
    \midrule
    &\textbf{RNNT-L} \\
    \midrule
    1  & 70.1 / 105.6 & 95.5 / 179.0 & 1.4/1.7 \\
    4  &  158.4 / 200.1 & 257.1 / 385.4 & 1.6/1.9 \\
    16 & 313.7 / 390.1 & 584.5 / 894.4 & 1.9/2.3 \\
    32 & 383.8 / 512.8 & 751.2 / 1403.6 & 2.0/2.7 \\
    \midrule
    &\textbf{RNNT-XXL} \\
    \midrule
    1 & 47.2 / 98.8 & 59.3 / 174.3 & 1.3/1.8 \\
    4 & 123.9 / 191.3 & 183.1 / 380.0 & 1.5/2.0  \\
    16 & 218.0 / 388.5 & 321.6 / 886.2  & 1.5/2.3 \\    
    32 & 248.4 / 532.7 & 352.2 / 1393.4 & 1.4/2.6 \\
    \bottomrule
    \end{tabular}
    \label{rnnt}
\end{table}

Table \ref{rnnt} presents our results with RNN Transducers.
The label-looping algorithm consistently speeds up decoding regardless of batch size. A larger relative speedup can be seen for larger batch sizes because, as the batch size grows, the baseline algorithm introduces more overhead since it is more likely that one utterance would need to wait for other utterances to advance $t$ in decoding. We also observe greater speed-up for non-encoder computations. Notably, for both models, our algorithm brings around 2.6X speed-up for batch-size=32 on non-encoder computations.

\subsection{Results with Token-and-Duration Transducers}

\begin{table}[t]
    \centering
    \caption{TDT decoding speed of different algorithms with different batch sizes on LibriSpeech test-other. Total RTFx / non-encoder RTFx. WER is 3.7\% for the L model, 2.8\% for XXL for our label-looping algorithm; for baseline, the WER fluctuates.}
    \begin{tabular}{c c c c}
    \toprule
    batch    & RTFx, baseline & RTFx, label-looping & rel. speedup  \\
    \midrule
    &\textbf{TDT-L} \\
    \midrule
    1 & 116.0 / 264.5 & 116.5 / 269.3 & 1.0 / 1.0 \\
    4 & 243.4 / 358.1 & 347.7 / 632.6 & 1.4 / 1.8 \\
    16 & 416.2 / 563.9 & 825.3 / 1615.9 & 2.0 / 2.9 \\
    32 & 477.2 / 696.7 & 1006.8 / 2634.2 & 2.1 / 3.8 \\
    \midrule
     &\textbf{TDT-XXL} \\
    \midrule
    1 & 66.1 / 241.3 & 66.1 / 251.7 & 1.0 / 1.0  \\
    4 & 147.9 / 258.8 & 219.5 / 584.5 & 1.5 / 2.3 \\
    16 & 243.7 / 478.3 & 374.8 / 1467.9 & 1.5 / 3.1\\
    32 & 265.2 / 616.3 & 394.0 / 2434.1 & 1.5 / 3.9 \\   
    \bottomrule
    \end{tabular}
    \label{tdt_results}
\end{table}

Label-looping experiments for the TDT model are shown in Table~\ref{tdt_results}. 
For the TDT baseline, we adopt the approximate method from~\cite{xu2023efficient}, which takes the minimum of predicted durations in the batch for advancement. Note that this method has non-deterministic outputs and is only included here for time comparison purposes.
Our algorithm is the first open-sourced implementation of exact and general-purpose TDT batched decoding. 
We observed that greater speed-ups can be seen with TDT models than conventional Transducers, and we see over 3.8X speed-up for non-encoder computation for both TDT-L and TDT-XXL models.

\section{Analysis}

We study the impact of precomputing the encoder and predictor projections, and compare the decoding RTFx with or without the precomputation. The results are in Table \ref{projection_study}. We see that overall, precomputing the encoder and predictor outputs brings over 20\% speedup for the decoding process, excluding encoder computations.

\begin{table}[t]
    \caption{Decoding RTFx (excluding encoder) between precomputation of projections and on-the-fly projections. Decoded on LibriSpeech test-other with RNNT-Large, bf16 precision.}
    \centering
    \begin{tabular}{c  c c c c}
    \toprule
         batch-size & w/o  & w/precomputation  & rel. speed-up \\
        \midrule
    1  & 147.7 &  179.0 & 1.21 \\
    4  & 303.8   &  385.4  & 1.27 \\
    32 & 1080.3   &  1403.6 & 1.30 \\    
    \bottomrule
    \end{tabular}
    \label{projection_study}
\end{table}

\begin{table}[t]
    \centering
    \caption{Performance on Conformer-Large models with 4x subsampling, bf16 precision. We report the total / non-encoder RTFx for decoding LibriSpeech test-other. WER is 3.7\%.}
    \begin{tabular}{ c c c c} 
    \toprule
     batch-size & baseline   &  ours & rel. speed-up\\
     \midrule
      1 & 48.2 / 63.2 & 79.5 / 130.8 & 1.6 / 2.1 \\
      4 & 104.7 / 123.1 & 192.9 / 262.3 & 1.8 / 2.1 \\
      16 & 187.2 / 243.1 & 341.6 / 577.5 & 1.8  / 2.4 \\
      32 & 226.0 / 328.7 & 399.0 / 866.2 & 1.8 / 2.6 \\
     \bottomrule
    \end{tabular}
    \label{4x}
\end{table}

In all experiments we report above, the encoder uses 8X subsampling. Another commonly used subsampling rate is 4X, and we test our proposed algorithm on a publicly available model\footnote{\href{https://hf.co/nvidia/stt_en_conformer_transducer_large}{\begin{scriptsize}\texttt{hf.co/nvidia/stt\_en\_conformer\_transducer\_large}\end{scriptsize}}} and show the results in Table \ref{4x}.
We see that with 4X models, our algorithm also significantly improves inference speed, up to 1.8X for total runtime and 2.6X for non-encoder computation.

It is worth mentioning that in our initial experiments, we also applied the beam search decoding algorithm to all RNNT models shown above and observed less than 0.1\% absolute WER improvement while observing a several times increase in decoding time. Thus, the accelerating greedy algorithm is crucial for the optimal speed vs accuracy trade-off and opens more possibilities for scaling the models to get the best performance.

\section{Combining Label-Looping with TorchScript and CUDA Graphs}

\begin{table}[t]
    \centering
    \caption{Combining Label-looping with TorchScript or CUDA graphs. Batch size = 32. Full-time RTFx / Non-encoder RTFx. Large RNNT and TDT models.}
    \begin{tabular}{c c c}
    \toprule
       model  & RNNT & TDT \\
       \midrule
        label-looping & 751.2 / 1403.6 & 1006.8 / 2634.2  \\
        + TorchScript & 882.1 / 1942.4 & 1118.0 / 3628.2 \\
        + CUDA graphs & 1232.7 / 5197.2 & 1393.4 / 9614.8 \\
    \bottomrule
    \end{tabular}
    \label{jit_cudagraph}
\end{table}

Since the speedup brought by our method is purely algorithmic, it can be combined with other methods that speed up the decoding, such as code compilation and GPU call optimization. In Table~\ref{jit_cudagraph} we report decoding results combining our algorithm with TorchScript\footnote{\url{https://pytorch.org/docs/stable/jit.html}} and with CUDA graphs~\cite{galvez24_speedoflight} methods. TorchScript and CUDA graph techniques aren't compatible, so, we do not include experiments combining them both.

Note that we show this table to demonstrate that our method is compatible with other methods. Because of the scope of this paper, this is the only table with results that combine those other optimizations. The speedup reported from all other tables is solely from the algorithmic changes of the label-looping algorithm. 

\section{Conclusion}
In this paper, we introduce a novel set of algorithms to improve the efficiency of Transducer models. The algorithm uses a label-looping design, which minimizes the number of prediction network calls, greatly reducing the runtime of decoding. Our experiments show that label-looping algorithms bring consistent speed-up over baseline batched decoding algorithms. In particular, we observe a speed-up of 2.0X for batch size = 32 for conventional Transducers and 2.1X for TDT models. 
Since our method is purely algorithmic, we show that it can be combined with other methods, such as TorchScript and CUDA graphs, to further accelerate Transducer decoding. 
Our algorithm is open-sourced through the NeMo~\cite{kuchaiev2019nemo} toolkit.

We will continue our research efforts to improve the algorithm for future work. In particular, since our algorithm greatly reduces the computational cost of the prediction network, this allows us to scale up the predictor for Transducer networks, e.g., using more layers or more sophisticated networks like Transformers.

\bibliographystyle{IEEEbib}
\bibliography{refs}

\end{document}